\documentstyle[tighten,preprint,eqsecnum,aps,epsf,floats]{revtex}  

\begin{document}

\preprint{\vbox{\hbox{JHU-TIPAC-99007}\hbox{KIAS--P99039}
\hbox{UTPT--99-09}\hbox{hep-ph/9907363}\hbox{July 1999}}}
 
\title{Phenomenology of $V_{ub}$ from Ratios of Inclusive $B$
Decay Rates}
\author{Junegone Chay,$\!^a$ Adam F. Falk,$\!^b$ Michael
Luke,$\!^c$ and Alexey A. Petrov$^b$\\ \medskip}\address{(a)
Department of Physics,  Korea University, Seoul 136-701, 
Korea\\ {\rm and}\\ Korea Institute for Advanced Study, Seoul
130-012, Korea\\ \medskip  (b) Department of Physics and
Astronomy, The Johns Hopkins University\\ 3400 North
Charles Street, Baltimore, Maryland 21218 U.S.A.\\ \medskip
(c) Department of Physics, University of Toronto\\ 60 St.~George
Street, Toronto, Ontario, Canada M5S 1A7}

\maketitle
\thispagestyle{empty}
\setcounter{page}{0}

\begin{abstract}%
We explore the theoretical feasibility of extracting $V_{ub}$
from two ratios built from $B$ meson inclusive partial decays,
$R_1=\Gamma(b\to u\bar cs')/3\Gamma(b\to cl\bar\nu)$ and
$R_2=[\Gamma(b\to cX)-\Gamma(b\to \bar cX)]/\Gamma(b\to
c\bar ud')$.  We discuss contributions to these quantities from
perturbative and nonperturbative physics, and show that they
can be computed with overall uncertainties at the level of 10\%.
\end{abstract}

\newpage

\section{Introduction}

The accurate measurement of $V_{ub}$ is one of the most
challenging theoretical and experimental problems in $B$
physics.  Its value is crucial for constraining the Unitarity
Triangle and probing the question of whether the CKM framework
is adequate for describing flavor physics in the standard
model.  The present best experimental values for $V_{ub}$, from
the inclusive decay $B\to X_ul\bar\nu$ and the exclusive process
$B\to\rho l\bar\nu$, are limited by model-dependence and other
theoretical errors.  New approaches to extracting $V_{ub}$ from
inclusive and exclusive semileptonic decays have been proposed
and are promising, but have not yet proven to be viable
experimentally.

In light of this situation, new methods for probing $V_{ub}$
are still needed.  In a recent paper~\cite{FaPe99}, we suggested
that it would be useful to attempt to measure the inclusive
production of ``wrong sign'' charm in $B$ decays, that is, to
look for evidence for the quark level transition $b\to u\bar
cs'$.  In particular, we proposed to study the ratio
$R_1=\Gamma(b\to u\bar cs')/ 3\Gamma(b\to cl\bar\nu)$, noting
that the theoretical expression for this quantity is in a
number of respects particularly well under control.  (Here
$s'$ and $d'$ are the flavor eigenstates, and we take
$m_s=m_d=0$.)  We went on to compute the leading perturbative
and nonperturbative corrections to the parton model result for
$R_1$.  The analysis of Ref.~\cite{FaPe99} also relied
implicitly on the use of parton-hadron duality.  This
assumption is common to all extractions of $V_{ub}$ from
inclusive $B$ decays, and while it is not unreasonable to expect
it to hold in this case, there is no rigorous proof that it
actually does.  Perhaps the near equality of the charged and
neutral $B$ meson lifetimes provides some empirical evidence
that duality is well respected in $B$ decays.

In this paper we will refine the analysis of Ref.~\cite{FaPe99}
in a number of respects.  First, we will include complete
radiative corrections to $R_1$ at next-to-leading order, that
is, all terms proportional to $\alpha_s(m_b)$ and
$\alpha_s^{n+1}\ln^n(m_W/m_b)$.  Second, we will include a set
of ``enhanced'' two loop terms, often referred to as ``BLM''
corrections~\cite{BLM83}, which are proportional to
$\alpha_s^2\beta_0$, where $\beta_0=11-2n_f/3$ is the first
coefficient in the QCD beta function.  It was pointed out
in Ref.~\cite{FaPe99} that these terms are not likely to be as
large in $R_1$ as in, for example $\Gamma(B\to X_cl\bar\nu)$,
because of the cancellation of the leading renormalon
ambiguity.  Indeed, our explicit calculation will show that
these terms contribute only at the level of ten percent. 
Third, in Ref.~\cite{FaPe99}, we also included the leading
nonperturbative contributions to the inclusive decay, which
come from annihilation processes and are proportional to
$16\pi^2 f_B^2/m_b^2$.  These terms are formally of order
$1/m_b^3$ but are enhanced by the two-body, rather than
three-body, phase space of the final state.  We found that in
charged $B$ decays, these processes can contribute at the order
of 5\%, while in neutral $B$ decays they turn out to be
negligible.  We will have little new to say about these
corrections, except that we will attempt to combine the
uncertainties from these contributions with those from the
radiative corrections to obtain an overall picture of the
reliability of the theoretical calculation.  We will also take
the opportunity to include an additional small ``hybrid''
contribution of order $\alpha_s\ln(m_W/m_b)\lambda_2/m_b^2$.

We will also propose that it is useful to consider a second
ratio, $R_2=[\Gamma(b\to cX)-\Gamma(b\to \bar cX)]/\Gamma(b\to
c\bar ud')$.  We will see that $R_2$ is theoretically clean in
a way which is similar to $R_1$, and we will extend all aspects
of our analysis of $R_1$ to include $R_2$.  The experimental
measurement of $R_2$ would certainly be challenging, but the
challenges would be distinct from those that confront the
measurement of $R_1$ and this second ratio deserves
separate consideration.

Finally, we will close by presenting an overall picture of the
theoretical understanding of $R_1$ and $R_2$, with our best
estimate of the remaining uncertainties and the future prospects
for reducing them.  We hope that this will provide an intriguing
goal for our experimental colleagues to aim for.

\section{$V_{\lowercase{ub}}$ from ratios of partial decay
rates}

In our previous paper~\cite{FaPe99}, we proposed that the quark
level process $b\to u\bar cs'$ would be a promising mode from
which to extract $V_{ub}$.  Final states with this combination
of quark flavors arise only from processes proportional to
$V_{ub}$, with no contributions from penguin diagrams or long
distance rescattering.  In the form of the ratio
\begin{equation}\label{R1def}
  R_1=\Gamma(b\to u\bar cs')/3\Gamma(b\to cl\bar\nu)\,,
\end{equation}
the theoretical expression is very well behaved.  The phase
space dependence on $x_c=m_c^2/m_b^2$ is identical in the
numerator and denominator, as are the leading nonperturbative
corrections of order $1/m_b^2$.  At tree level and in the limit
$m_b\to\infty$, then, we have simply
\begin{equation}
  R_1=|V_{ub}/V_{cb}|^2\,\left\{1+O(\alpha_s,1/m_b^2)\right\}\,.
\end{equation}
In Ref.~\cite{FaPe99}, we first included radiative corrections
at leading logarithmic order, which has the effect of
multiplying the expression for $R_1$ by a factor
$\chi\simeq1.09$.  We also computed the leading radiative
correction to the decay processes $b\to u\bar cs'$ and $b\to
cl\bar\nu$, which although formally subleading is numerically
substantial.  The result was an expression of the form
\begin{equation}\label{gdef}
  R_1=\chi\,|V_{ub}/V_{cb}|^2\,\left\{
  1+g(x_c){\alpha_s\over\pi}+\ldots\right\}\,.
\end{equation}
With $x_c=0.09$ and $\alpha_s=\alpha_s(m_b)=0.22$, the one loop
radiative correction is $g\alpha_s/\pi=0.21$, so indeed it
is large and should be included.  However, the scale $\mu$ at
which $\alpha_s$ ought to be evaluated was not fixed by our
calculation, leading to a significant remaining uncertainty. 
This can be resolved only with a full next-to-leading order
calculation, which we will perform in the next section.  We
will find that the partial calculation of Ref.~\cite{FaPe99}
was correct to within approximately 20\%.

Unfortunately, the experimental measurement of $R_1$ is
extremely challenging.  The largest background to observing the
quark level process $b\to u\bar cs'$ is $b\to c\bar cs'$, the
rate for which is approximately a factor of 100 larger.  The
measurement is made more difficult by the fact that the only
quantity which is well predicted theoretically is the ratio of
fully {\it inclusive\/} rates, while many of the experimental
techniques for rejecting $b\to c\bar cs'$ involve tagging on a
particular hadronic final state.  Although the measurement of
$R_1$ may be feasible, it will hardly be straightforward. 
Nevertheless, relevant experimental techniques already are being
developed~\cite{Mar99}, and the excellent capability of the
BaBar and BELLE detectors to vertex individually the boosted
$B$ mesons also will improve the prospects for this
measurement~\cite{Kino99}.

There is another ratio which one might consider, which avoids
the necessity of rejecting the $b\to c\bar cs'$ background.  Let
$\Gamma(b\to cX)=\Gamma(b\to c\bar ud')+\Gamma(b\to c\bar
cs')$ be the fully inclusive production of $c$ in nonleptonic
$b$ decays, and $\Gamma(b\to \bar cX)=\Gamma(b\to u\bar
cs')+\Gamma(b\to c\bar cs')$ be the inclusive production of
$\bar c$.  Also, define $\Gamma(b\to cX')=\Gamma(b\to c\bar
ud')$, that is, the inclusive production of $c$ {\it
without\/} an accompanying $\bar c$.  Note that 
$\Gamma(b\to c\bar cs')<\Gamma(b\to c\bar ud')$, so the
measurement of $\Gamma(b\to cX')$ does not require rejecting an
overwhelming background.  Then let
\begin{equation}\label{R2def}
  R_2={\Gamma(b\to cX)-\Gamma(b\to\bar cX)\over\Gamma(b\to cX')}
  \equiv 1-\delta_2\,.
\end{equation}
We see that $\Gamma(b\to c\bar cs')$ cancels in the numerator
of $R_2$.  In terms of quark level transitions,
\begin{equation}
  \delta_2={\Gamma(b\to u\bar cs')\over
  \Gamma(b\to c\bar ud')}\,.
\end{equation}
At tree level, $\delta_2=|V_{ub}/V_{cb}|^2$.  Unlike in $R_1$,
there is no leading logarithmic correction to $\delta_2$, since
these contribute identically to the numerator and the
denominator.  The leading radiative correction arises at order
$\alpha_s(m_b)$, and the leading nonperturbative corrections at
order $\alpha_s\ln(m_W/m_b)\lambda_2/m_b^2$ and
$16\pi^2f_B^2/m_b^2$.

The experimental advantage of $R_2$ is that the large $b\to
c\bar cs'$ background cancels.  The difficulty is that in
order for $R_2$ to be sensitive to $|V_{ub}/V_{cb}|^2$, 
both $\Gamma(b\to cX)$ and $\Gamma(b\to cX')$ must be measured
with an accuracy of better than 1\%.  This may prove to be as
challenging as rejecting $b\to c\bar cs'$, or even more so, but
it involves a distinct set of problems.  It will be up to the
experimental community to determine whether this measurement is
feasible or not.

A quantity which would be more attractive experimentally is
\begin{equation}\label{R3def}
  R_3={\Gamma(b\to cX)-\Gamma(b\to\bar cX)
  \over\Gamma(b\to cl\bar\nu)}\,,
\end{equation}
since doing so avoids the requirement of determining
$\Gamma(b\to cX')$ precisely.  Unfortunately, $R_3$ cannot be
computed with the very small uncertainties of $R_1$ and $R_2$. 
This is simply because, unlike $R_1$ and $R_2$, the
calculation of $R_3$ requires that the ratio  $\Gamma(b\to
cX')/\Gamma(b\to cl\bar\nu)=R_1/\delta_2$ be known with a
theoretical accuracy of better than 1\%.  This is well beyond
the level of precision presently attainable, due to
the perturbative and nonperturbative contributions which we
will discuss below.  Thus, we will not consider $R_3$ further. 
However, we do note that in anticipation of theoretical
advances, it may well be useful for experimentalists to measure
$R_3$ as well as $R_1$ and $R_2$.  Alternatively, if
$\Gamma(b\to cX')/\Gamma(b\to cl\bar\nu)$ could itself be {\it
measured\/} with the required precision, then it could be used
to construct $R_2$ from the experimentally more accessible
ratio $R_3$.

\section{Radiative corrections at next-to-leading order}

In Ref.~\cite{FaPe99}, we included two subsets of radiative
corrections.  First, we used an effective Lagrangian evaluated
at the scale $\mu=m_b$,
\begin{equation}\label{leff}
  {\cal L} = -\frac{4G_F}{\sqrt{2}} \,V_{ub} \,\Bigl[
  C_1(\mu) \bar u_\alpha \gamma_\mu P_Lb_\alpha 
  \,\bar s'_\beta \gamma^\mu P_L c_\beta
  +C_2(\mu)\bar u_\alpha \gamma_\mu P_L b_\beta 
  \,\bar s'_\beta \gamma^\mu P_L c_\alpha\Bigr]+{\rm h.c.}\,.
\end{equation}
Employing such a Lagrangian has the effect of summing all
logarithms of the form $\alpha_s^n\ln^n(m_W/m_b)$.  At leading
log order, $C_1(m_b)\simeq 1.11$ and
$C_2(m_b)\simeq -0.24$~\cite{Leff}.  Second, we performed a
partial one-loop calculation of the radiative correction to the
decay rate itself.  Although we included the largest
contributions, our calculation was incomplete because we 
omitted terms which were not proportional to
$(C_1+C_2/N_c)^2$.  This allowed us to consider only gluon
exchanges within color-singlet currents, simplifying the result
enormously.  At leading log order terms of this form contribute
96\% of the total decay rate, so we hoped that the error from
this approximation would not be too large.  The combined
radiative correction which we found, keeping all terms of order
$\alpha_s^n\ln^n(m_W/m_b)$ and of order $\alpha_s$ but only
{\it some\/} of order $\alpha_s^2\ln(m_W/m_b)$, was
$\chi[1+g(x_c)\alpha_s/\pi]$, where $\chi$ came from QCD
running between $M_W$ and $m_b$ and $g(x_c)$ was a finite
radiative correction to the decay rate. For $x_c=0.09$ and
$\alpha_s=\alpha_s(m_b)=0.21$, this gave 1.32, or a combined
correction of about 30\%. 

The largest ambiguity in this result comes from the scale $\mu$
at which the one-loop correction to the decay process is to be
evaluated.  This ambiguity can be removed only by performing a
full calculation at next-to-leading log order, including
consistently all terms of order $\alpha_s^{n+1}\ln^n(m_W/m_b)$. 
We present the results of such a calculation here.  We have
followed closely the analogous calculation of $b\to c\bar ud'$
by Bagan {\it et al.}~\cite{BBBG94}, and have used their partial
results where appropriate.

We refer the reader to the excellent exposition of
Ref.~\cite{BBBG94} for a detailed discussion of the method of
the analysis, and here present only our results.  A brief
synopsis of our calculation is found in the Appendix.  We
write the answer in the form
\begin{eqnarray}
  R_1=|V_{ub}/V_{cb}|^2\,\left\{
  1+r_1(\mu,x_c)+\dots\right\}\,,
  \nonumber\\
  \delta_2=|V_{ub}/V_{cb}|^2\,\left\{
  1+r_2(\mu,x_c)+\dots\right\}\,,
\end{eqnarray}
where $r_i(\mu,x_c)$ are of order $\alpha_s$.  Here $\mu$ is the
renormalization scale and $x_c=m_c^2/m_b^2$ is the ratio of the
heavy quark pole masses.  We take $m_q=0$ for $q=u,d,s$, which
is in this process an excellent approximation even for the
strange quark.

\begin{figure}[htb]
\begin{center}
\mbox{\epsfxsize 3in
\epsfbox{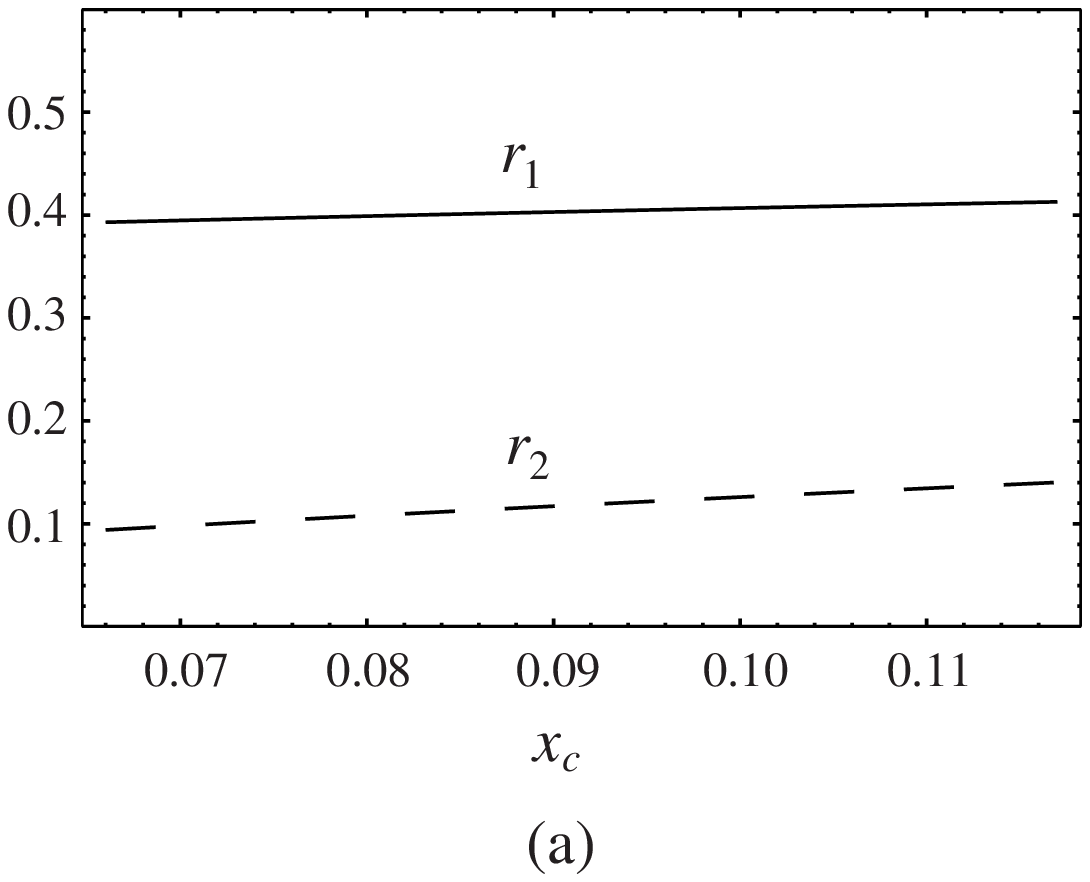}\hskip0.5cm \epsfxsize 3in
\epsfbox{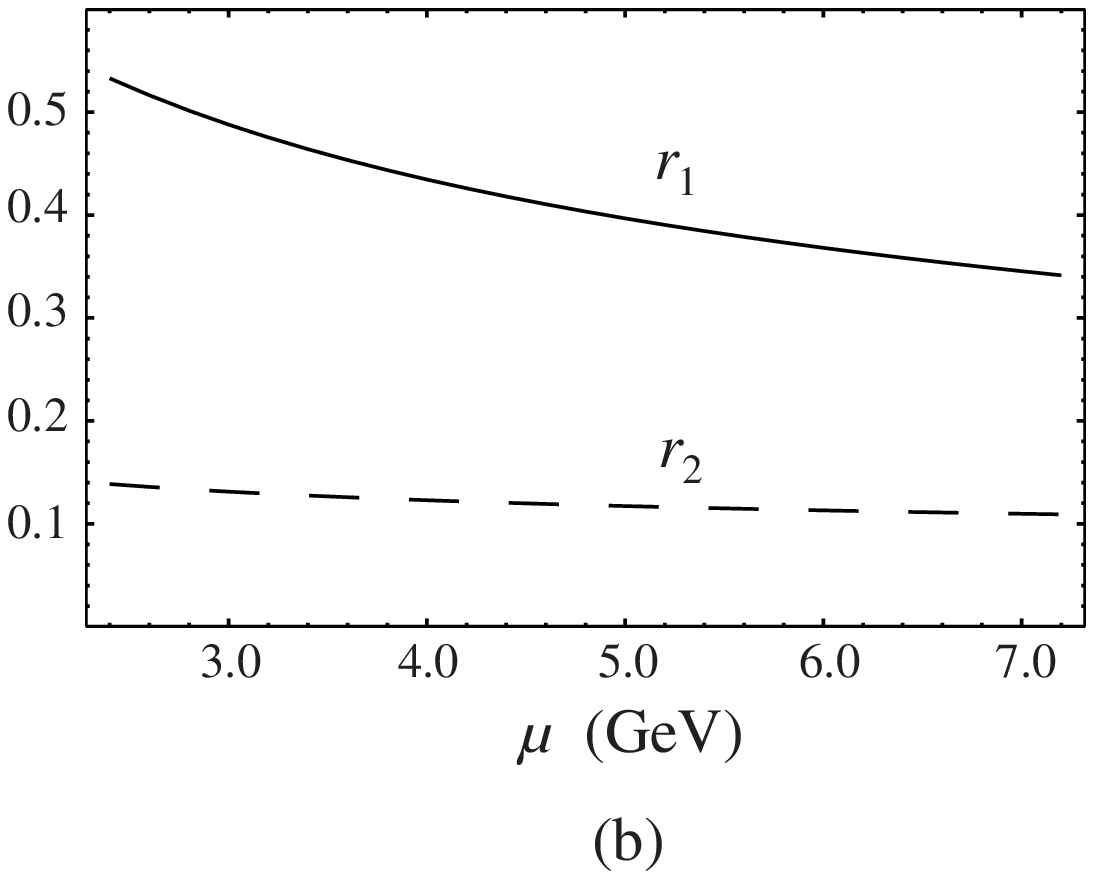}}
\end{center}
\caption{{Variation of $r_i$ with $x_c$ and $\mu$. (a)
$r_i(x_c)$ for $4.5\,{\rm GeV}\le m_b\le 5.1\,{\rm GeV}$ and
$\mu=m_b$.  (b)~$r_i(\mu)$ for $2.4\,{\rm GeV}\le \mu\le
7.2\,{\rm GeV}$ and
$m_b=4.8$\,GeV.}}
\label{radplots}
\end{figure}

Taking the reference values $m_b=4.80\,$GeV, $m_c=1.45\,$GeV
(so $x_c=0.09$), and $\mu=4.8\,$GeV, we find
\begin{equation}
  r_1(\mu,x_c)=0.40\qquad{\rm and}\qquad
  r_2(\mu,x_c)=0.12\,.
\end{equation}
Because the leading logarithms cancel in the ratio $\delta_2$,
$r_2$ starts at order $\alpha_s$ rather than at order
$\alpha_s\ln(M_W/m_b)$ and is considerably smaller than $r_1$.
In Fig.~\ref{radplots}, we display the variation of $r_i$ with
$x_c$ and
$\mu$.
In Fig.~\ref{radplots}a, we vary $m_b$ between $4.5\,$GeV and
$5.1\,$GeV, fixing $m_c$ by the heavy quark symmetry constraint
$m_b-m_c=3.35\,$GeV and taking $\mu=m_b$.  We see that the
dependence on $x_c$ is very mild; over the conservative range
considered, $r_1$ varies by $\pm0.01$ and $r_2$ by $\pm0.02$. 
In Fig.~\ref{radplots}b, we fix $m_b=4.8\,$GeV and vary $\mu$
between ${1\over2}m_b$ and ${3\over2}m_b$.  For $r_2$, the
dependence is soft, approximately $\pm0.02$.  However, for
$r_1$ a significant $\mu$-dependence remains even at
next-to-leading order.  For $\mu$ as low as ${1\over2}m_b$, we
have $r_1=0.53$.  We will choose to assign an asymmetrical
error of $(+0.10,-0.05)$ to the
$\mu$-dependence of $r_1$.  Combining the variation in $x_c$
and $\mu$, then, we find the results
\begin{equation}\label{radresult}
  r_1=0.40^{+0.10}_{-0.05}\qquad{\rm and}\qquad
  r_2=0.12\pm0.03\,.
\end{equation}
In the partial calculation of our previous paper, we found
$\chi(1+g\alpha_s/\pi)=1.32$, to be compared with $1+r_1$ here. 
We now see that this approximation underestimated the correct
next-to-leading order result by 0.08, or 20\%.  While our
incomplete treatment gave a reasonable result, including the
full calculation at this order turns out to be important.

\section{BLM Corrections}

At the next order in $\alpha_s$, a consistent leading-log
calculation requires the three-loop anomalous dimensions of the
operators in ${\cal L}$ and the two-loop matrix elements. 
However, since the effects of the running are not large, a
useful estimate of these corrections is obtained by simply
taking the two-loop matrix element of the singlet operator;
this corresponds to neglecting terms of order $\alpha_s^3
\ln(m_W/m_b)$ relative to $\alpha_s^2$.  A further
simplification is obtained by only retaining the so-called
``BLM'' two-loop corrections~\cite{BLM83}, which are enhanced
by a factor of $\beta_0\equiv 11-2 N_f/3$, where $N_f=3$ is the
number of light quark flavors.\footnote{Note that since the
size of the BLM correction has nothing to do with the scale
$\mu$ we used in the previous section to evaluate the
coefficients in the effective Lagrangian, we only interpret the
BLM correction as an estimate of the full two-loop matrix
elements, not as providing information on the scale at which the
one-loop corrections should be evaluated.  We also do not
include charm among the light quarks at the scale $\mu=m_b$.}
This class of two-loop corrections is computed easily by
performing a weighted  integral over the one-loop result
calculated with a gluon mass~\cite{SV94}. While the BLM
corrections are not formally dominant in any limit of QCD, in
many processes they are found empirically to be the largest
part of  the two-loop term.  In this section we calculate the
BLM corrections to
$R_1$ and $\delta_2$.

\begin{figure}[htb]
\begin{center}
\vskip-1cm
\mbox{\hskip-0.25in\epsfxsize 6.5in
\epsfbox{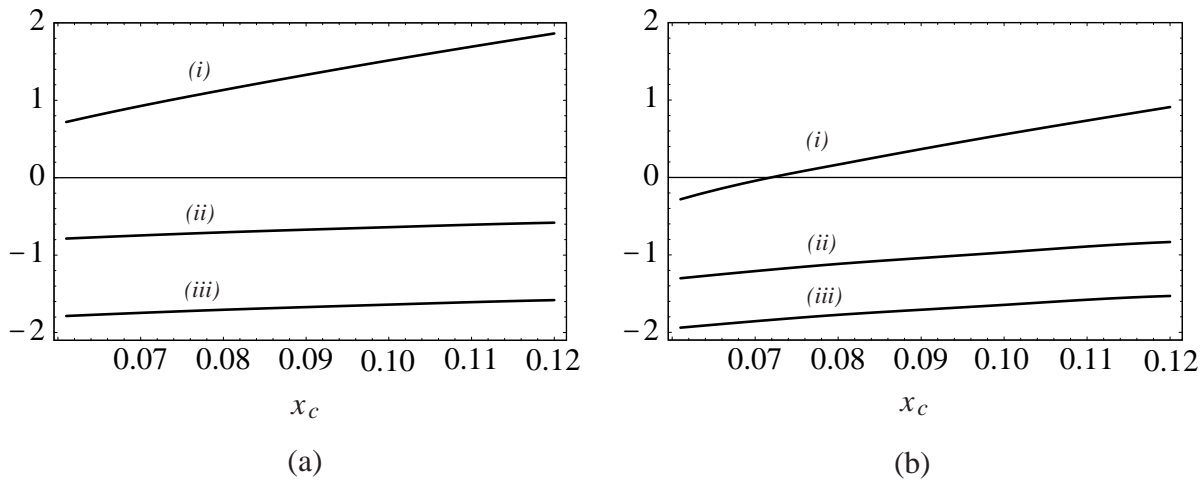}}
\end{center}
\caption{(a) The one-loop coefficient  $a_1^X$ and (b) the
BLM-enhanced two-loop coefficient $a_2^X$ to ({\it i})
$b\rightarrow u\bar cs$, ({\it ii}) $b\rightarrow c\bar u d$
and  ({\it iii}) $b\rightarrow c l\bar\nu$ decays.}
\label{aplots}
\end{figure}

These ratios require the BLM corrections to $b\rightarrow u\bar
cs^\prime$, $b\rightarrow c\bar u d^\prime$ and $b\rightarrow
c l\bar \nu$. The calculation of the BLM corrections to
$b\rightarrow u\bar c s^\prime$ is identical to that for
$b\rightarrow c\bar c s^\prime$~\cite{LLSS97} with one of the
charm quark masses set to zero, so we refer the reader to
Ref.~\cite{LLSS97} for details.  The corrections are
particularly simple because the corrections to the $bu$ and
$\bar cs^\prime$ currents factorize.   This feature also allows
the BLM corrections to $b\rightarrow c\bar ud^\prime$ to be
extracted easily from existing calculations of the semileptonic
decay rate and $R(e^+e^-\rightarrow \mbox {hadrons})$. 
Finally, the BLM corrections to semileptonic
$b\rightarrow c$ decays were calculated in Ref.~\cite{LSW95}. 
Writing each decay rate as
\begin{equation}
  \Gamma(b\rightarrow X)\propto 1+a_1^X\,
  {\alpha_s(m_b)\over\pi}+
  a_2^X\,\beta_0\left({\alpha_s(m_b)\over\pi}\right)^2+\dots,
\end{equation}
we plot the $a_i^X$'s in Fig.~\ref{aplots} for each of the
relevant decays.

\begin{figure}[htb]
\begin{center}
\mbox{\epsfxsize 4.0in\epsfbox{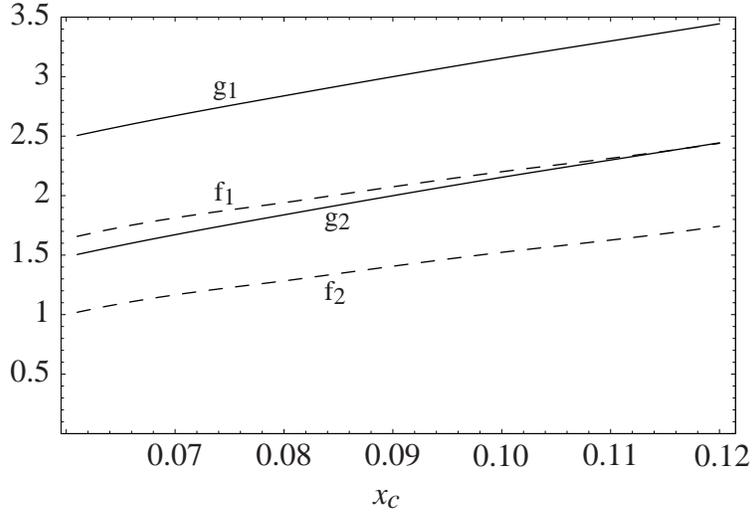}}
\end{center}
\caption{The one-loop coefficients $g_i$ (solid) and
BLM-enhanced two-loop coefficients $f_i$ (dashed).}
\label{blmcorrections}
\end{figure}

For the purpose of evaluating the quality of the perturbation
series for the matrix elements, one should compare this two loop
result to the one loop correction $g(x_c)$ defined in
Eq.~(\ref{gdef}).  The reason is that neither the BLM correction
nor $g(x_c)$ has a logarithmic dependence on $M_W$.  Hence we
neglect for the moment the full NLO corrections of the previous
section and write
\begin{eqnarray}
  R_1=\chi \left|{V_{ub}\over V_{cb}}\right|^2\left\{1+g_1(x_c)
  {\alpha_s(m_b)\over\pi}+
  f_1(x_c)\beta_0\left({\alpha_s(m_b)\over\pi}\right)^2
  +\dots\right\}\,,\nonumber\\
  \delta_2=\left|{V_{ub}\over V_{cb}}\right|^2\left\{1+g_2(x_c)
  {\alpha_s(m_b)\over\pi}+
  f_2(x_c)\beta_0\left({\alpha_s(m_b)\over\pi}\right)^2
  +\dots\right\}\,.
\end{eqnarray}
The coefficients $g_i(x_c)$ and $f_i(x_c)$ are plotted in
Fig.~\ref{blmcorrections}.
Taking $x_c=0.09$ and $\mu=m_b$, we
find for $R_1$ the results
\begin{eqnarray}
  &&g_1(0.09)\,{\alpha_s(m_b)\over\pi}
  =3.0\,{\alpha_s(m_b)\over\pi}=0.21\,,\nonumber\\
  &&f_1(0.09)\,\beta_0\left({\alpha_s(m_b)\over\pi}\right)^2
  =2.1\,\beta_0\left({\alpha_s(m_b)\over\pi}\right)^2
  =0.09\,,
\end{eqnarray}
and for $\delta_2$,
\begin{eqnarray}
  &&g_2(0.09)\,{\alpha_s(m_b)\over\pi}
  =2.0\,{\alpha_s(m_b)\over\pi}=0.14\,,\nonumber\\
  &&f_2(0.09)\,\beta_0\left({\alpha_s(m_b)\over\pi}\right)^2
  =1.4\,\beta_0\left({\alpha_s(m_b)\over\pi}\right)^2
  =0.06\,.
\end{eqnarray}
While it is not formally consistent to include these
corrections in the NLO calculations of the previous section, we
can  use them to shift the central values of the $r_i$'s,
\begin{eqnarray}
  r_1(\mu=m_b, x_c=0.09)&\rightarrow& r_1+
  1.09\times f_1(0.09)=0.40+0.10=0.50\,,\nonumber\\
  r_2(\mu=m_b, x_c=0.09)&\rightarrow& r_2+
  f_2(0.09)=0.12+0.06=0.18\,.
\end{eqnarray}
Note that both corrections are somewhat larger than
the error estimates from varying the renormalization scale in
the previous section.  Varying $m_b$ between 4.5~GeV and
5.1~GeV yields an additional uncertainty, which we estimate to
be $(+0.005,-0.025)$ on $r_1$ and $\pm0.015$ on $r_2$.

\section{Nonperturbative corrections}

In addition to the perturbative corrections discussed so far,
there are also nonperturbative contributions to $R_1$ and
$\delta_2$ which are sensitive to the configuration of the
initial $B$ meson.  As discussed in Ref.~\cite{FaPe99}, the
cancellation of the tree level phase space factor eliminates
terms in $R_1$ of order $\Lambda_{\rm QCD}/m_b$.
Furthermore, while there could be in principle subleading terms
at order $1/m_b^2$ proportional to the HQET parameters
$\lambda_1$ and $\lambda_2$~\cite{lambda}, these cancel in the
ratio as well.

This can be understood by examining the possible
sources  of dependence on the charm quark mass.  These
corrections enter  the calculations of $b\to u\bar cs'$ and
$b\to c\bar ud'$ either from kinematic phase space functions
or from performing spin sums.  Consider the Fierzed form of the
effective Lagrangian (\ref{leff}),
\begin{equation}\label{leff2}
  {\cal L} = -\frac{4G_F}{\sqrt{2}} \,V_{ub} \,\Bigl[
  C_1(\mu)\bar s'_\beta \gamma_\mu P_Lb_\alpha 
  \,\bar u_\alpha  \gamma^\mu P_L c_\beta
  +C_2(\mu)\bar s'_\beta \gamma_\mu P_L b_\beta 
  \,\bar s'_\beta\bar u_\alpha \gamma^\mu P_L
  c_\alpha\Bigr]+{\rm h.c.}\,.
\end{equation}
Substituting $s'\to d'$ and $V_{ub}\to V_{cb}$, and exchanging
$u$ and $c$, we obtain the Lagrangian responsible for $b\to
c\bar ud'$.  Recall that we take $m_d=m_s=0$.  The calculation
of the decay $b\to c\bar ud'$ involves the computation of the
polarization tensor with a massive $c$ quark and a massless
$\bar u$ antiquark, whereas the $b\to u\bar cs'$ decay involves
a massless $u$  and a massive $\bar c$. Hence the Fierz
transformation by itself does not guarantee the cancellation of
all terms in the ratios $R_i$. However, because of the $VV-AA$
structure of the polarization tensor, and the fact that $VV$
and $AA$ correlation functions are symmetric under $m_u
\leftrightarrow m_c$~\cite{Re80}, the total decay rate is also
symmetric under interchange of the quark masses.  Thus, both the
phase space and $1/m_b^2$ corrections cancel in $\delta_2$.  To
extend the argument to $R_1$, note that in the Fierzed form and
at tree level, the decay $b\to c\bar u d'$ is the same as $b\to
u\tau\bar\nu_\tau$, for which the nonperturbative corrections
were computed in Ref.~\cite{Fa94}.  This discussion elaborates
the observation made in Ref.~\cite{FaPe99} that these
corrections cancel.  A general argument that the terms
proportional to $\lambda_1$ (but not to $\lambda_2$) cancel in
all ratios of $B$ decays was given in Ref.~\cite{Bigi92}.

There remain, however, mixed terms proportional to
$\alpha_s\lambda_2/m_b^2$, which need not cancel in the
ratios.  Recall that $\lambda_2=\langle B|\,\bar
b_vg\sigma\cdot Gb_v\,|B\rangle/12M_B$, where $b_v$ is the
effective $b$ field of HQET, is related to the hyperfine
interaction of the $b$ quark chromomagnetic moment with the
light degrees of freedom in the $B$ meson~\cite{lambda,Bigi92}. 
The chromomagnetic operator is obtained by attaching a gluon to
one of the quarks in the final state, before the operator product
expansion is performed.  In the effective theory defined by the
Lagrangian (\ref{leff}), terms of order $\alpha_s\lambda_2$
come from two sources.  First, they arise from one-loop
radiative corrections to the operator product expansion itself,
in which case they are quite small.  Second, the color
structure allows terms proportional to $C_1C_2$ to arise at
tree level from the interference of the color singlet and color
exchanged operators.  These terms are really of order
$\alpha_s\ln(M_W/m_b)\lambda_2/m_b^2$ and hence are enhanced
over the others.  They were first calculated for the decay $b\to
c\bar ud'$ in Ref.~\cite{Bigi92}.

In fact, a straightforward argument shows that these terms are
equal in size and of the opposite sign in $b\to u\bar cs'$ as
compared to $b\to c\bar ud'$.  Consider again the Fierzed form
of the effective Lagrangian (\ref{leff2}).  Because of the
color structure when the operators are written in this form, the
term proportional to $C_1C_2\lambda_2$ is generated by attaching
a gluon to the $\bar cu$ loop, performing the operator product
expansion, and extracting the chromomagnetic moment operator
$\bar b\sigma^{\mu\nu}G_{\mu\nu}b$.  This term is odd
under the exchange $u\to c$ and $\bar c\to \bar u$, or
equivalently $m_u^2\leftrightarrow m_c^2$, which can be seen
immediately by inspection of the relevant Feynman diagrams. 
Alternatively, simply note that the quark and antiquark
produced by the left-handed current carry opposite magnetic
moments.  To obtain the result for $b\to u\bar cs'$, we take the
limit $m_u=0$; as noted above, for $b\to c\bar ud'$, we can take
$m_c=0$ and $m_u\to m_c$.  Since the general result is odd in
$(m_c^2-m_u^2)$, the two limiting results are the negatives of
each other, as promised.

\begin{figure}[htb]
\begin{center}
\vskip-1cm
\mbox{\epsfxsize 3in
\epsfbox{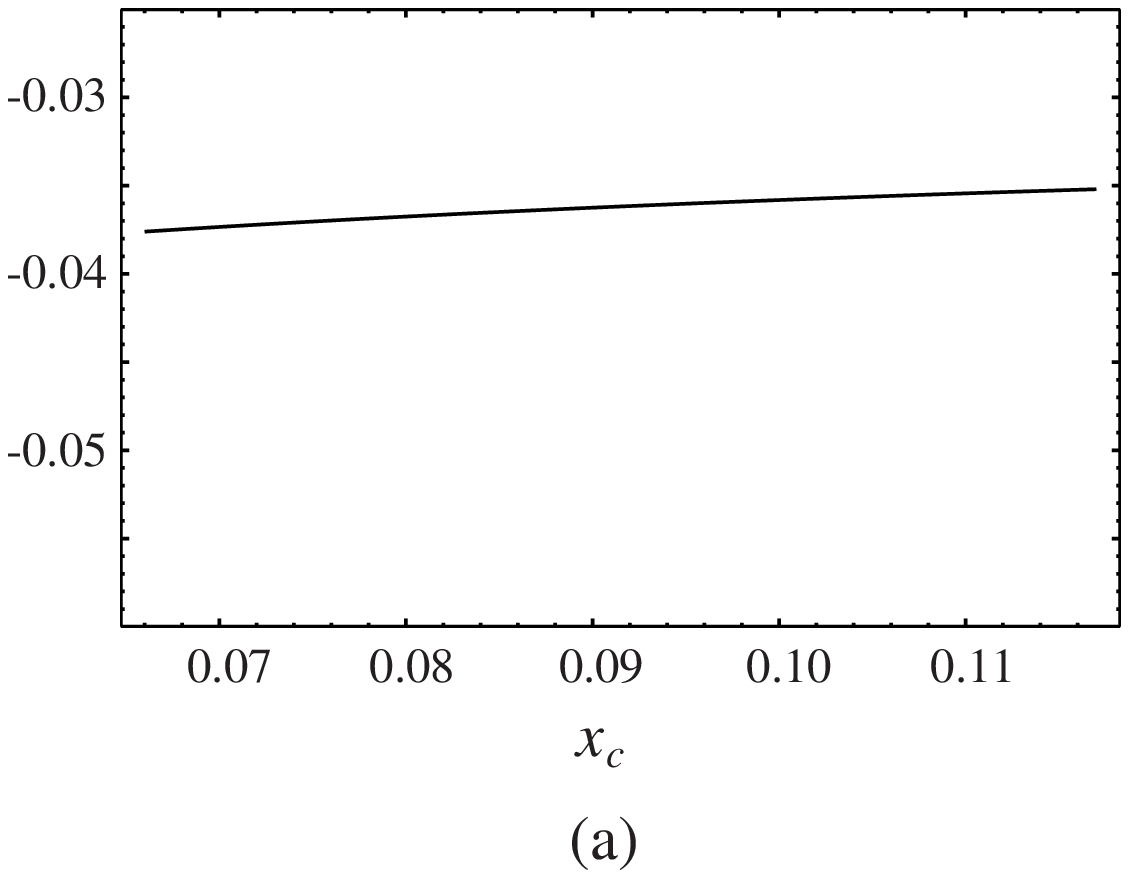}\hskip0.5cm \epsfxsize 3in
\epsfbox{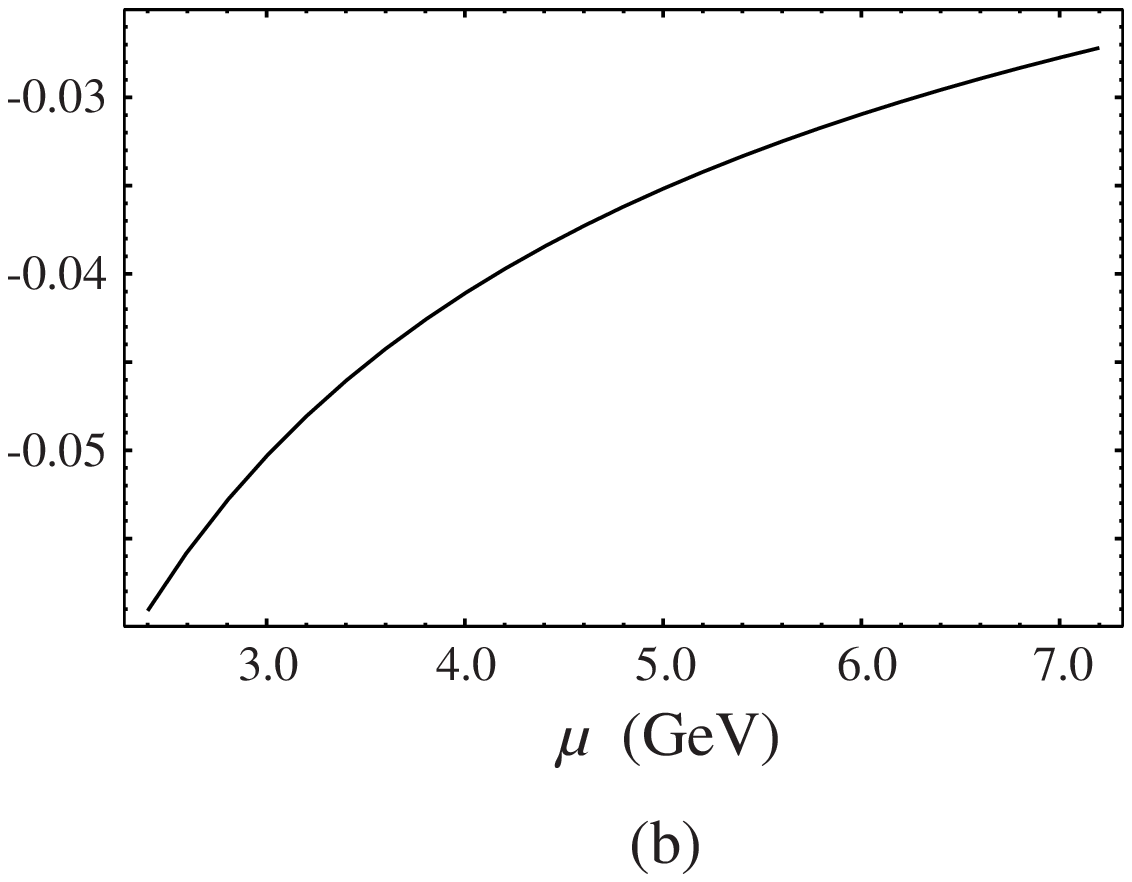}}
\end{center}
\caption{Variation of $\ell_2$ with $x_c$ and $\mu$. (a)
$\ell_2(x_c)$ for $4.5\,{\rm GeV}\le m_b\le 5.1\,{\rm GeV}$ and
$\mu=m_b$.  (b)~$\ell_2(\mu)$ for $2.4\,{\rm GeV}\le \mu\le
7.2\,{\rm GeV}$ and $m_b=4.8$\,GeV.}
\label{l2plots}
\end{figure}

We write the result as fractional corrections to $R_1$ and
$\delta_2$ (suppressing for the moment the radiative
corrections of the previous sections),
\begin{eqnarray}
  R_1&=&|V_{ub}/V_{cb}|^2\,\Big\{1+\ell_2(\mu,x_c)+\dots
  \Big\}\,,
  \nonumber\\
  \delta_2&=&|V_{ub}/V_{cb}|^2\,\Big\{
  1+2\ell_2(\mu,x_c)+\dots\Big\}\,,
\end{eqnarray}
where
\begin{equation}
  \ell_2(\mu,x_c)=C_1(\mu)C_2(\mu)\,{16(1-x_c)^3\over f(x_c)}\,
  {\lambda_2(\mu)\over m_b^2}
\end{equation}
and $f(x_c)=1-8x_c+8x_c^3-x_c^4-12x_c^2\ln x_c$ is the tree
level phase space function.  For consistency, $C_1(\mu)$ and
$C_2(\mu)$ should be evaluated at leading order, to include
only terms of order
$\alpha_s^n\ln^n(M_W/m_b)\lambda_2/m_b^2$.  The scale
dependence of $\lambda_2$ is given by~\cite{lam2}
\begin{equation}
  \lambda_2(\mu)=\lambda_2(m_b)\Big[\alpha_s(\mu)/\alpha_s(m_b)
  \Big]^{9/25}\,.
\end{equation}
For $\mu=m_b=4.8\,$GeV, and with $\lambda_2(m_b)=0.12\,$GeV
fixed by the $B-B^*$ mass splitting~\cite{lam2}, we find
$\ell_2=-0.036$.  The variation of $\ell_2$ with $x_c$ and $\mu$
is shown in Fig.~\ref{l2plots}.
We see that, as with the
radiative corrections, the only significant uncertainty comes
from the choice of renormalization scale.  We assign the tiny
error $\pm0.001$ to $\ell_2$ from the variation with $x_c$, and
the larger asymmetrical error $(+0.010,-0.015)$ to the variation
with $\mu$.  Combining these in quadrature, we see that the
first is negligible, and we estimate
\begin{equation}
  \ell_2=-0.036^{+0.010}_{-0.015}\,.
\end{equation}
Noting that the fractional correction to $\delta_2$ is $2\ell_2$
and comparing to the radiative correction $r_2$
(\ref{radresult}), we see that this is actually an important
uncertainty for $\delta_2$.

The leading purely nonperturbative corrections to $R_1$ and
$\delta_2$ arise at order $1/m_b^3$.  The largest such
corrections are associated with annihilation processes such as
$b\bar u\to \bar cs'$, since they are enhanced by a relative
phase space factor of $16\pi^2$.  They are also spectator
dependent, contributing differently to $B^-$ and $\overline
B{}^0$ decays.  In Ref.~\cite{FaPe99}, we discussed the
derivation and computed the annihilation terms in $R_1$.  Here
we will recall those results, as well as present results for
$\delta_2$.  As before, we will present a fractional
correction, of the form
\begin{eqnarray}
  R_1&=&|V_{ub}/V_{cb}|^2\,\Big\{1+a_1(B^-,
  \overline B{}^0)+\dots\Big\}\,,
  \nonumber\\
  \delta_2&=&|V_{ub}/V_{cb}|^2\,\Big\{
  1+a_2(B^-,\overline B{}^0)+\dots\Big\}\,,
\end{eqnarray}
where all other corrections have been momentarily suppressed.
The terms $a_i$ depend on nonperturbative matrix elements of
four-quark operators, parameterized by ``bag factors'' $B_i$
and $\epsilon_i$~\cite{NeSa97}:
\begin{eqnarray}\label{eBdefs}
  \langle B_q|\,
  \bar b_L\gamma^\mu q_L\,\bar q_L\gamma_\mu b_L
  \,|B_q\rangle&=&
  \textstyle{1\over4}\,f_{B_q}^2m_{B_q}^2\,B_1\,,\nonumber\\
  \langle B_q|\,
  \bar b_Rq_L\,\bar q_L b_R
  \,|B_q\rangle&=&
  \textstyle{1\over4}\,f_{B_q}^2m_{B_q}^2\,B_2\,,\nonumber\\
  \langle B_q|\,
  \bar b_L\gamma^\mu T^aq_L\,\bar q_L\gamma_\mu T^ab_L
  \,|B_q\rangle&=&
  \textstyle{1\over4}\,f_{B_q}^2m_{B_q}^2\,
  \epsilon_1\,,\nonumber\\
  \langle B_q|\,
  \bar b_RT^aq_L\,\bar q_L T^ab_R
  \,|B_q\rangle&=&
  \textstyle{1\over4}\,f_{B_q}^2m_{B_q}^2\,
  \epsilon_2\,.
\end{eqnarray}
In the vacuum insertion ansatz, only color single operators
contribute to the decay and we have $B_1=B_2=1$ and
$\epsilon_1=\epsilon_2=0$.  More generally, the color octet
parameters $\epsilon_i$ are of order $1/N_c$ in the limit
$N_c\to\infty$.

In terms of these parameters, we find the corrections
\begin{eqnarray}
  a_1(B^-)&=&{16\pi^2f_B^2(1-x_c)^2\over m_b^2f(x_c)}\,
  \Big\{(C_1+\textstyle{1\over3}C_2)^2
  \big[(1+2x_c)B_2-(1+\textstyle{1\over2}x_c)B_1\big]
  \nonumber\\ &&\qquad\qquad\qquad\qquad 
  \mbox{}+\textstyle{2\over3}C_2^2\big[(1+2x_c)\epsilon_2
  -(1+\textstyle{1\over2}x_c)\epsilon_1\big]\Big\}\,,\nonumber\\
  a_1(\overline B{}^0)&=&{16\pi^2f_B^2(1-x_c)^2\over
  m_b^2f(x_c)}\,\sin^2\theta_C\,
  \Big\{(C_2+\textstyle{1\over3}C_1)^2
  \big[(1+2x_c)B_2-(1+\textstyle{1\over2}x_c)B_1\big]
  \nonumber\\ &&\qquad\qquad\qquad\qquad 
  \mbox{}+\textstyle{2\over3}C_1^2\big[(1+2x_c)\epsilon_2
  -(1+\textstyle{1\over2}x_c)\epsilon_1\big]\Big\}
\end{eqnarray}
for $R_1$, and
\begin{eqnarray}
  a_2(B^-)&=&{16\pi^2f_B^2(1-x_c)^2\over m_b^2f(x_c)}\,
  \Big\{(C_1+\textstyle{1\over3}C_2)^2
  \big[(1+2x_c)B_2-(1+\textstyle{1\over2}x_c)B_1\big]
  \nonumber\\ &&\qquad\qquad\qquad\qquad 
  \mbox{}+\textstyle{2\over3}C_2^2\big[(1+2x_c)\epsilon_2
  -(1+\textstyle{1\over2}x_c)\epsilon_1\big]
  \nonumber\\ &&\qquad\qquad\qquad\qquad
  -\textstyle{1\over3}(C_1^2+6C_1C_2+C_2^2)B_1
  -2(C_1^2+C_2^2)\epsilon_1\Big\}\,,\nonumber\\
  a_2(\overline B{}^0)&=&-{16\pi^2f_B^2(1-x_c)^2\over
  m_b^2f(x_c)}\,\cos2\theta_C\,
  \Big\{(C_2+\textstyle{1\over3}C_1)^2
  \big[(1+2x_c)B_2-(1+\textstyle{1\over2}x_c)B_1\big]
  \nonumber\\ &&\qquad\qquad\qquad\qquad\qquad\qquad 
  \mbox{}+\textstyle{2\over3}C_1^2\big[(1+2x_c)\epsilon_2
  -(1+\textstyle{1\over2}x_c)\epsilon_1\big]\Big\}
\end{eqnarray}
for $\delta_2$.  In deriving the corrections to the denominator
of $\delta_2$, we have adapted the results of Ref.~\cite{NeSa97}
for the channel
$b\to c\bar ud'$.  With $m_b=\mu=4.8\,$GeV,
$f_B=200\,$MeV and $\sin\theta_C=0.22$, we find
\begin{eqnarray}\label{aresults}
  a_1(B^-)&=&-0.48B_1+0.54B_2
  -0.018\epsilon_1+0.021\epsilon_2\,,\nonumber\\
  a_1(\overline B{}^0)&=&-0.00034B_1+0.00038B_2
  -0.018\epsilon_1+0.020\epsilon_2\,,\nonumber\\
  a_2(B^-)&=&-0.43B_1+0.54B_2
  -1.14\epsilon_1+0.021\epsilon_2\,,\nonumber\\
  a_2(\overline B{}^0)&=&0.0063B_1-0.0071B_2
  +0.34\epsilon_1-0.38\epsilon_2\,.
\end{eqnarray}

Note that in $a_2(B^-,\overline B{}^0)$, the coefficients of
$\epsilon_i$ are quite large.  This reflects the potentially
significant contribution of color-octet annihilation processes
to nonleptonic $B$ decays, as observed in Ref.~\cite{NeSa97}. 
Since the parameters $\epsilon_i$ are not known well, the large
size of these terms introduces a problematic uncertainty into
the denominator of $\delta_2$.  If we take the vacuum insertion
ansatz, in which $\epsilon_i$ do not contribute, we have
\begin{eqnarray}
  a_1(B^-)&=&0.062\,,\qquad\qquad
  a_1(\overline B{}^0)=4.4\times10^{-5}\,,\nonumber\\
  a_2(B^-)&=&0.112\,,\qquad\qquad
  a_2(\overline B{}^0)=-8.1\times10^{-4}\,.
\end{eqnarray}
Unfortunately, it is hard to assess the uncertainty due to
nonzero $\epsilon_i$.  For want of a better procedure, let us
survey briefly the available models for estimating the
matrix elements (\ref{eBdefs}).  The most reliable of these,
in principle, is the lattice QCD result~\cite{UKQCD98}
\begin{eqnarray}
  B_1(m_b) &=& 1.06 \pm 0.08\,,\qquad\qquad\
  \epsilon_1(m_b) = -0.01 \pm 0.03\,,\nonumber\\
  B_2(m_b) &=& 1.01 \pm 0.06\,,\qquad\qquad\
  \epsilon_2(m_b) = -0.02 \pm 0.02\,,
\end{eqnarray}
where the quoted errors include neither quenching errors nor
the systematic uncertainty due to the extrapolation to the
chiral limit.  Both of these issues can be addressed in future,
more precise calculations.  There also exist calculations in
the  framework of QCD sum rules, which give~\cite{Cher95}
\begin{eqnarray}
  &&B_1 \simeq 1\,,\qquad\qquad\epsilon_1 \simeq -0.15\,,
  \nonumber \\ 
  &&B_2 \simeq 1\,,\qquad\qquad\epsilon_2 \simeq 0\,,
\end{eqnarray}
and~\cite{Ba98}
\begin{eqnarray}
  B_1(m_b) &=& 1.01 \pm 0.01\,,\qquad\qquad\
  \epsilon_1(m_b) = -0.08 \pm 0.02\,,\nonumber\\
  B_2(m_b) &=& 0.99 \pm 0.01\,,\qquad\qquad\
  \epsilon_2(m_b) = -0.01 \pm 0.03\,,
\end{eqnarray}
as well as an HQET QCD sum rule calculation 
which yields~\cite{ChYa99}
\begin{eqnarray}
  B_1(m_b) &=& 0.96 \pm 0.04\,,\qquad\qquad\
  \epsilon_1(m_b) = -0.14 \pm 0.01\,,\nonumber\\
  B_2(m_b) &=& 0.95 \pm 0.02\,,\qquad\qquad\
  \epsilon_2(m_b) = -0.08 \pm 0.01\,.
\end{eqnarray}
While the $B_i$'s are consistently within 5\% or so of unity,
there is a large spread in the values of the $\epsilon_i$
parameters.  Of course, it makes no sense to ``average
over models'' as a method for assigning values to $B_i$ and
$\epsilon_i$.  Instead, we adopt the procedure of taking
the lattice QCD results as central values but inflating the
errors both to be conservative and to reflect the variety of
values which QCD sum rules yield for $\epsilon_i$.  To be even
more conservative, we inflate the errors {\it
symmetrically,} so that we use the sum rules results to set the
magnitude, but not the sign, of the uncertainty in the lattice
calculations.  The central values and errors which we choose
are then
\begin{eqnarray}\label{bepsnums}
  B_1(m_b) &=& 1.06 \pm 0.10\,,\qquad\qquad\
  \epsilon_1(m_b) = -0.01 \pm 0.10\,,\nonumber\\
  B_2(m_b) &=& 1.01 \pm 0.10\,,\qquad\qquad\
  \epsilon_2(m_b) = -0.02 \pm 0.10\,,
\end{eqnarray}
One could imagine a less conservative procedure, especially if
one had particular confidence in one of the calculations quoted
above, but this is not the approach which we will follow.

Inserting these parameters (\ref{bepsnums}) into the solution
(\ref{aresults}) for the nonperturbative corrections, we find
\begin{eqnarray}
  a_1(B^-)&=&0.04\pm0.07\,,\qquad\qquad
  a_1(\overline B{}^0)=0\pm0.003\,,\nonumber\\
  a_2(B^-)&=&0.10\pm0.13\,,\qquad\qquad
  a_2(\overline B{}^0)=0\pm0.05\,.
\end{eqnarray}
Note that the fractional error associated with the weak
annihilation contribution is particularly large for $\delta_2$
in charged $B$ decays, due to its enhanced sensitivity to
$\epsilon_i$.

\section{Phenomenology of $V_{\lowercase{ub}}$ from $R_1$ and
$R_2$}

Combining the results of the previous sections, we now present
estimates for central values and uncertainties for $R_1$ and
$\delta_2 = 1- R_2$.  Due to the sizable flavor-dependent
corrections associated with the spectator contributions, we
present our results separately for charged and neutral $B$
decays, which we denote by introducing a suitable superscript. 
Our results take the form
\begin{eqnarray}
  R_1^- &=& \left| V_{ub}/V_{cb} \right|^2
  \left [ 1 + r_1(x_c, \mu) + {\ell}_2 (x_c, \mu) + a_1 (B^-)
  \right ],
  \nonumber \\
  R_1^0 &=& \left| V_{ub}/V_{cb} \right|^2
  \left [ 1 + r_1(x_c, \mu) + {\ell}_2 (x_c, \mu) + a_1
  (\overline B{}^0)\right ],
  \nonumber \\
  \delta_2^- &=& \left| V_{ub}/V_{cb} \right|^2
  \left [ 1 + r_2(x_c, \mu) + 2 {\ell}_2 (x_c, \mu) + a_2 (B^-)
  \right ],
  \nonumber \\
  \delta_2^0 &=& \left| V_{ub}/V_{cb} \right|^2
  \left [ 1 + r_2(x_c, \mu) + 2 {\ell}_2 (x_c, \mu) + a_2
  (\overline B{}^0)\right ],
\end{eqnarray}
where $r_i$ comes from perturbative QCD radiative corrections
and $\ell_2$ and $a_i$ represent fractional corrections due to
nonperturbative effects. Putting our results together, we find
\begin{eqnarray} \label{err}
  R_1^- &=& \left| V_{ub}/V_{cb} \right|^2
  \left [ 1.50 ^{+0.10}_{-0.05}
  {}^{+0.005}_{-0.025}{}^{+0.010}_{-0.015}
  \pm 0.07
  \right ],
  \nonumber \\
  R_1^0 &=& \left| V_{ub}/V_{cb} \right|^2
  \left [ 1.46^{+0.10}_{-0.05}
  {}^{+0.005}_{-0.025}{}^{+0.010}_{-0.015}
  \pm 0.005
  \right ],
  \nonumber \\
  \delta_2^- &=& \left| V_{ub}/V_{cb} \right|^2
  \left [ 1.21 \pm 0.03
  \pm0.015
  {}^{+0.015}_{-0.030} \pm 0.13
  \right ],
  \nonumber \\
  \delta_2^0 &=& \left| V_{ub}/V_{cb} \right|^2
  \left [ 1.11 \pm 0.03
  \pm0.015
  {}^{+0.015}_{-0.030} \pm 0.05
  \right ].
\end{eqnarray}
In these expressions, the first error is our estimate of the
uncertainty from NLO perturbative QCD corrections, second is
due to uncertainty in the BLM part of the two-loop corrections,
the third represents uncertainty in the $\alpha_s \ln (M_W/m_b)
\lambda_2/m_b^2$ term, and fourth is due to spectator-dependent
$1/m_b^3$ effects. We expect the net effect of other $1/m_b^3$
effects, not enhanced by the phase space factor of $16
\pi^2$, to be safely below the level of the estimated
uncertainty.

We can combine the errors quoted in Eq.~(\ref{err}) by taking
into account the correlations among the various sources of
theoretical uncertainty.  However, we would like to emphasize
that this procedure of estimating and combining theoretical
errors, while widespread, is purely conventional and has no
rigorous statistical meaning.  With this in mind, our best
estimate of  the central values and overall uncertainties is
\begin{eqnarray}
  &&R_1^- = \left| V_{ub}/V_{cb} \right|^2
  \Big[1.50\pm0.15\Big]\,,\qquad\qquad
  \delta_2^- = \left| V_{ub}/V_{cb} \right|^2
  \Big[1.21\pm0.15\Big]\,,\nonumber\\
  &&R_1^0 = \left| V_{ub}/V_{cb} \right|^2
  \Big[1.46\pm0.10\Big]\,,\qquad\qquad\,
  \delta_2^0 = \left|V_{ub}/V_{cb} \right|^2
  \Big[1.11\pm0.10\Big]\,.
\end{eqnarray}

Although the uncertainties are generally smaller for
the neutral $B$ decays, in all cases the theoretical errors
are at approximately the level of ten percent. For $R_1$, the
error is dominated by residual uncertainties in the
next-to-leading order radiative corrections.  Reducing them
substantially would require no less than a
next-to-next-to-leading order calculation.  For
$\delta_2$, the uncertainties come primarily from the poorly
known strong matrix elements needed for the annihilation
contributions, especially from the color octet bag factors
$\epsilon_i$.  The best prospect for improvement here is in a
future generation of unquenched lattice calculations.  Were
these to become available, overall theoretical errors in
$\delta_2$ at the five percent level would be within reach.

In summary, we have studied the possibility of
extracting the CKM matrix element $V_{ub}$ from
ratios of inclusive nonleptonic $B$ decays. We have
shown that the ratios of inclusive decay widths defined by
Eqs.~(\ref{R1def}) and (\ref{R2def}) are impressively ``clean''
theoretically. We have estimated the impact and
uncertainty associated with the NLO radiative corrections, and
have included the BLM part of the two-loop term.
In addition, we have studied the impact of the leading
non-perturbative corrections on $R_1$ and $R_2$.
There is no doubt that the measurement of the $R_1$ and $R_2$
would be a challenging enterprise.  Unfortunately, the somewhat
experimentally easier ratio of $R_3$ of Eq.~(\ref{R3def}) has
larger theoretical uncertainties, from radiative corrections
which would need to be computed at better than the 1\% level
before the method could be used to extract $V_{ub}$. 
Nonetheless, the ratios $R_1$ and $R_2$ of nonleptonic decay
widths offer a new and tantalizing approach to measuring the
important but poorly known CKM matrix element $V_{ub}$.

\acknowledgements
We are indebted to P.~Ball for allowing us to incorporate into
our calculation her computer code for part of the radiative
corrections at next-to-leading order.  Support for J.C.\ was
provided by the Korea Ministry of Education under Grant BSRI
98-2408, by  the German-Korean Scientific Exchange Program
DFG-446-KOR-113/72/0, and by the KOSEF/NSF Scholar Exchange
Program.  Support for A.F.\ and A.P.\ was provided by the United
States National Science Foundation under Grant PHY--9404057 and
National Young Investigator Award PHY--9457916, and by the
United States Department of Energy under Outstanding Junior
Investigator Award DE--FG02--94ER40869.  A.F.\ also is
supported by the Alfred P.\ Sloan Foundation and is a Cottrell
Scholar of the Research Corporation.  Support for M.L.\ was
provided by the Natural Sciences and Engineering Research
Council of Canada  and the Alfred P.\ Sloan Foundation.

\appendix
\section{}

Here we outline the procedure for calculating radiative
corrections to $R_1$ and $\delta_2$ with next-to-leading log
(NLO) accuracy.  This calculation amounts to the computation
of  perturbative corrections to $b \to u \bar c s'$. We follow
closely the procedure outlined in Ref.~\cite{BBBG94}. 

\begin{figure}[htb]
\begin{center}
\mbox{\epsfxsize=4.5in\epsfbox{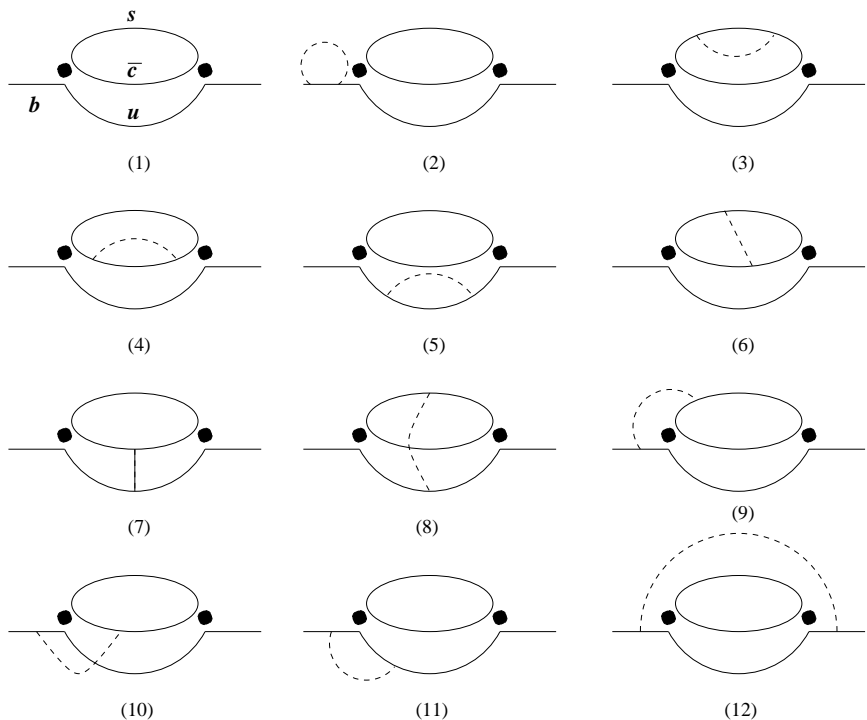}}
\end{center}
\caption{\label{diagrams}Feynman diagrams for the calculation 
of QCD corrections to $b \to u \bar c s'$. Black dots represent 
insertions of the operator $O_1$ or $O_2$.  Dashed lines
represent gluons.}
\end{figure}

A full NLO calculation of QCD radiative corrections to $b \to u
\bar c s'$  involves the computation of the eleven loop diagrams
depicted in Fig.~\ref{diagrams}. These diagrams can be 
conveniently ``packaged'' into five classes. In fact, only the
diagrams of a single class will have to be computed, as the
others may be extracted from existing calculations of
perturbative QCD corrections to polarization tensors and
inclusive semileptonic $b$ decays.  As discussed in
Ref.~\cite{BBBG94}, this is achieved by the application of Fierz
relations to $O_1=\bar u_\alpha \gamma_\mu P_Lb_\alpha\,\bar
s'_\beta \gamma^\mu P_L c_\beta$ and $O_2=\bar u_\alpha \gamma_\mu P_L b_\beta
\,\bar s'_\beta \gamma^\mu P_L c_\alpha$ (note that our choice
of $O_1$ and $O_2$ is opposite to that of Ref.~\cite{BBBG94}). 
Although in general Fierz symmetry is broken by regularization,
Na\"\i ve Dimensional Regularization (NDR) along with the choice
of evanescent operators advocated in Ref.~\cite{BurWe90}
actually preserves Fierz relations for renormalized operators. 
This allows us to use the Fierz transformation on a diagram by
diagram basis~\cite{BBBG94}, which significantly simplifies the
task of computing the graphs of Fig.~\ref{diagrams} in the
limit of massless $u$ and $s$ quarks.

We use the effective Lagrangian of Eq.~(\ref{leff}) calculated
at NLO accuracy.  It is convenient to define the
Wilson coefficients $C_\pm(\mu)$ of the multiplicatively
renormalized operators $O_\pm = {1\over2}(O_1\pm O_2)$.  At this
order, they are given by
\begin{eqnarray}
  C_\pm(\mu) = C_1(\mu) \pm C_2(\mu) &=&
  L_\pm(\mu)
  \left[
  1 + \frac{\alpha_s(m_W) - \alpha_s(\mu)}{4\pi}\,
  \frac{\gamma^{(0)}_\pm}{2 \beta_0}
  \left(
  \frac{\gamma^{(1)}_\pm}{\gamma^{(0)}_\pm}-
  \frac{\beta_1}{\beta_0}
  \right)+
   \frac{\alpha_s(m_W)}{4\pi}\, B_\pm
  \right] 
  \nonumber \\
  &=& L_\pm(\mu)
  \left[
  1 + \frac{\alpha_s(m_W) - \alpha_s(\mu)}{4\pi}
  \,R_\pm + \frac{\alpha_s(\mu)}{4\pi}\, B_\pm
  \right],
\end{eqnarray}
where for $N_f$ flavors and $N_c=3$ colors the anomalous
dimensions $\gamma_\pm =\gamma^{(0)}_\pm (\alpha_s/4\pi) + 
\gamma^{(1)}_\pm (\alpha_s/4\pi)^2 + ~...$ are $\gamma^{(0)}_+
= 4$, $\gamma^{(0)}_- = -8$, $\gamma^{(1)}_+ = -7 + 4 N_f/9$,
and $\gamma^{(1)}_- = -14 - 8N_f/9$.  The QCD $\beta$ function
is given by
\begin{eqnarray}
  \beta &=& - g_s \left[\beta_0 \frac{\alpha_s}{4 \pi} + 
  \beta_1 \left(\frac{\alpha_s}{4\pi}\right)^2 + \dots\,
  \right],
  \nonumber \\
  \beta_0 &=& 11 - \frac{2}{3} N_f\,, \qquad
  \beta_1 = 102 - \frac{38}{3} N_f\,.
\end{eqnarray}
The coefficients of $O_\pm$ at leading logarithmic order are
\begin{equation}
  L_\pm (\mu) = \left(\frac{\alpha_s(m_W)}{\alpha_s(\mu)}
  \right)^{{\gamma^{(0)}_\pm}/{2 \beta_0}}.
\end{equation}
Finally, the result of two loop matching at $\mu = m_W$
for the effective Lagrangian (in NDR) is contained in 
\begin{equation} \label{bees}
  B_\pm = \pm B\, \frac{N_c \mp 1}{2 N_c}\,, \qquad{\rm
  where}\qquad B = 11.
\end{equation} 
As advocated in Ref.~\cite{BBBG94}, it is useful to
combine Eq.~(\ref{bees}) with the anomalous dimensions,
\begin{equation}
  R_\pm = B_\pm + 
  \frac{\gamma^{(0)}_\pm}{2 \beta_0}
  \left(
  \frac{\gamma^{(1)}_\pm}{\gamma^{(0)}_\pm}-
  \frac{\beta_1}{\beta_0}
  \right),
\end{equation}
so that $R_\pm$ are independent of the renormalization
scheme.  Following Ref.~\cite{BBBG94}, the decay rate for $b \to
u\bar c s'$ can be expressed as
\begin{eqnarray}
  \Gamma (b \to u \bar c s') &=&
  {G_F^2m_b^5|V_{ub}|^2\over192\pi^3} ~\Biggl \{
  2 L_+^2(\mu) + L_-^2(\mu) 
  + \frac{\alpha_s(m_W) -
  \alpha_s(\mu)}{\pi}
  \left[2 L_+^2(\mu)R_+ + L_-^2(\mu)R_-\right]
  \nonumber
  \\
  &&\qquad\quad\mbox{}+
  \frac{\alpha_s(\mu)}{2\pi} \left[
  \left[L_+(\mu) + L_-(\mu)\right]^2
  c_{22}(x_c) + \left[L_+(\mu) - L_-(\mu)\right]^2 c_{11}(x_c)
  \right]
  \nonumber \\
  &&\qquad\quad\mbox{}+ \frac{\alpha_s(\mu)}{3\pi} 
  \left[L_+(\mu)^2 - L_-^2(\mu)\right] c_{12}(x_c)
  \Biggr \}\,,
\end{eqnarray}
where we define
$c_{11}(x_c) = G_c + G_d$, $c_{22}(x_c) = G_a + G_b$, and
$c_{12}(x_c) = G_a + G_b +G_e + B$. Here the five classes
of graphs $G_i$ are defined as
\begin{eqnarray} \label{gs}
  \frac{\alpha_s}{\pi} G_a &=&
  K ~Im \Big[ (2) + (2)^\dagger + (3) + (9) + (9)^\dagger + (12)
  \Big],
  \nonumber \\
  \frac{\alpha_s}{\pi} G_b &=&
  K ~Im \Big[ (4) + (5) + (7) \Big],
  \nonumber \\
  \frac{\alpha_s}{\pi} G_c &=&
  K ~Im \Big[ (2) + (2)^\dagger + (5) + (11) + (11)^\dagger +
  (12)\Big],
  \nonumber 
  \\
  \frac{\alpha_s}{\pi} G_d &=&
  K ~Im \Big[ (3) + (4) + (6) \Big],
  \nonumber \\
  \frac{\alpha_s}{\pi} G_e &=&
  K ~Im \Big[ (4) + (8) + (10) + (10)^\dagger + (11) +
  (11)^\dagger\Big],
\end{eqnarray}
with $K=192 \pi^3/m_b^6f(x_c)$ and $f(x_c)
=1-8x_c+8x_c^3-x_c^4-12x_c^2\ln x_c$.  Examination of
Eq.~(\ref{gs}) reveals that  $G_a = G_c$ and can be extracted
from the calculation of  QCD corrections to the semileptonic
decay $b \to u \tau \bar \nu_\tau$~\cite{CJK95},
\begin{eqnarray}
  G_a(x_c)&=&\frac{1}{48f(x_c)} \Bigl\{
  4(1-x_c)(75-539x_c-476x_c^2+18x_c^3)
  \nonumber \\
  &&\qquad\qquad\mbox{}-
  16\pi^2(3-24x_c-36x_c^2+16x_c^3-2x_c^4)-
  3456x_c^2\Big(\zeta(3)-{\rm Li}_3(x_c)\Big)
  \nonumber \\
  &&\qquad\qquad\mbox{}-
  96(1-8x_c+36x_c^2+16x_c^3-2x_c^4)\,{\rm Li}_2(x_c)
  \nonumber
  \\ 
  &&\qquad\qquad\mbox{}-  
  8(1-x_c^2)(31-320x_c+31x_c^2)\ln(1-x_c)
  \nonumber \\ 
  &&\qquad\qquad\mbox{}-48
  \Bigl(2x_c+15x_c^2-\textstyle{94\over3}x_c^3
  +\textstyle{31\over6}x_c^4
  -8\pi^2x_c^2+24x_c^2\,{\rm Li}_2(x_c)
  \nonumber \\ 
  &&\qquad\qquad\qquad\quad\mbox{}+
  2(1-x_c^2)(1-8x_c+x_c^2)\ln(1-x_c)\Bigr)\ln x_c\Bigr\}\,.
\end{eqnarray}
\begin{figure}
\centerline{\epsfxsize=3.5in\epsfbox{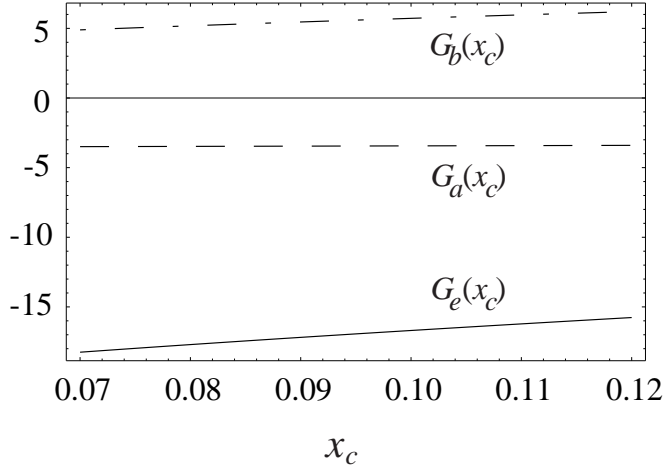}}
\caption{\label{depend}Variation of $G_{\{a,b,e\}}$ with $x_c$.
The dashed line is $G_a(x_c)$, the dash-dotted line is 
$G_b(x_c)$, and the solid line is $G_e(x_c)$.}
\end{figure} 
Likewise, $G_b = G_d$ and can be extracted from the computation of 
perturbative QCD corrections to the polarization operator 
of vector and axial currents~\cite{Re80,Vol95}, 
\begin{eqnarray}
  G_b(x_c)&=& \frac{1}{48 f(x_c)} \Bigl\{
  (1-x_c)(18-476x_c-539x_c^2+75x_c^3)
  \nonumber \\
  &&\qquad\qquad\mbox{}+
  4x_c^2(36+8x_c-x_c^2)(\pi^2-3\ln^2 x_c)
  \nonumber \\
  &&\qquad\qquad\mbox{}-
  2(1-x_c^2)\Big(31-320x_c+31x_c^2
  -12(1-8x_c+x_c^2)\ln x_c\Big)\ln(1-x_c)
  \nonumber
  \\
  &&\qquad\qquad\mbox{}-
  2x_c(132+90x_c-308x_c^2+31x_c^3)\ln x_c
  \nonumber \\
  &&\qquad\qquad\mbox{}+24\bigl(2-16x_c-36x_c^2+8x_c^3-x_c^4
  +
  12x_c^2\ln x_c\bigr)\,{\rm Li}_2(x_c)\nonumber \\
  &&\qquad\qquad\mbox{}
  +864x_c^2\Big(\zeta(3)-{\rm Li}_3(x_c)\Big)  \Bigr \}\,.
\end{eqnarray}
Hence $G_e$ is the only new quantity which needs to be computed.
We have done this numerically, modifying a subroutine given to
us by P.~Ball~\cite{BBBG94}.  Since each of $G_a, G_b, G_c$, and
$G_d$ is related to a physical process, the individual classes
of the diagrams are gauge invariant and free of infrared 
singularities.  We present  plots of $G_i(x_c)$ in
Fig.~\ref{depend}, with $\mu=m_b$ and $m_b-m_c=3.35\,$GeV. We
see that the dependence on the quark masses is not very strong.

To obtain the radiative correction to $R_1$, the results of
this calculation must be combined with the one-loop result for
$b\to c\ell\bar\nu$~\cite{Nir89}.  For the denominator of
$\delta_2$, we use the result of Ref.~\cite{BBBG94}.  The
radiative corrections to the two observables are plotted in
Fig.~\ref{radplots} of the text.

\tighten

\end{document}